\documentclass[a4paper,fleqn,12pt]{cas-sc}

\usepackage{textcomp}
\usepackage[numbers]{natbib}
\usepackage{amsmath}
\usepackage{chemformula}

\newcommand{\highlight}[1]{{#1}}

\ExplSyntaxOn
\cs_set:Npn \__first_footerline: {}
\ExplSyntaxOff

\begin{document}
\let\WriteBookmarks\relax
\def\floatpagepagefraction{1}
\def\textpagefraction{.001}

\shorttitle{Spectroscopy of Er$^{3+}$ in K$_2$YF$_5$ }    

\shortauthors{P.\ S.\ Solanki et.\ al.}  

\title [mode = title]{Spectroscopy and Crystal-Field Analysis of Low -Symmetry Er$^{3+}$ Centres in K$_2$YF$_5$ Microparticles}

\author[1,2]{Pratik S. Solanki}[]
\credit{Investigation, Formal analysis, Writing - original draft}
\author[1,2]{Michael F. Reid}[orcid=0000-0002-2984-9951]
\cormark[1]
\credit{Conceptualization, Software, Investigation, Supervision, Writing - review and editing}
\ead{mike.reid@canterbury.ac.nz}
\author[1,2]{Jon-Paul R. Wells}[orcid=0000-0002-8421-6604]
\cormark[1]
\credit{Supervision, Writing - review and editing, Visualization, Resources, Funding Acquisition, Project Administration}
\ead{jon-paul.wells@canterbury.ac.nz}

\affiliation[1]{organization={The School of Physical and Chemical Sciences, University of Canterbury},
            addressline={PB4800 Christchurch 8140, New Zealand}
           }

\affiliation[2]{organization={The Dodd-Walls Centre for Photonic and Quantum Technologies},
            addressline={New Zealand}
           }
        
\cortext[1]{Corresponding authors}

\fntext[1]{}


\begin{abstract}
  K$_2$YF$_5$ crystals doped with lanthanide ions have a variety of possible optical applications. Owing to the low symmetry of the system, the crystal structure cannot be unambiguously determined by x-ray diffraction. However, electron-paramagnetic resonance studies have demonstrated that lanthanide ions substitute for yttrium in sites of C$_{\rm s}$ local symmetry. In this work, we use high-resolution absorption and laser spectroscopy to determine electronic energy levels for  Er$^{3+}$ ions in K$_2$YF$_5$ microparticles. A total of 39 crystal-field energy levels, distributed among 7 multiplets of the Er$^{3+}$ ion, have been assigned. 
This optical data is used for crystal-field modelling of the electronic structure of Er$^{3+}$ in K$_2$YF$_5$. Our model is fitted not only to the electronic energy levels, but also to the ground-state g-tensor. This magnetic-splitting data defines the axis system of the calculation, avoiding ambiguities associated with low-symmetry crystal-field fits.  
\end{abstract}

  


 

\maketitle

\section{Introduction}

Lanthanide ion doped potassium yttrium pentafluoride crystals (K$_2$YF$_5$) are promising optically active media that have been investigated for a variety of applications, including phosphors, thermoluminescence dosimeters, laser crystals, and  passive Q-switching \cite{kui2006thermoluminescence,faria2004thermoluminescence,khaidukov2021study,wang2019synthesis,Loiko2016}.
An understanding of the electronic structure of lanthanide ions in this host requires a knowledge of the crystal structure.  X-ray diffraction is unable to distinguish between two possible space groups, $Pnam$ and $Pna2_1$.
However, electron paramagnetic resonance (EPR) studies indicate that the structure is  monoclinic, with space group $Pnam$ \cite{loncke2007epr,zverev2011electron}. The unit cell dimensions are $a=10.820$\,\AA, $b=6.613$\,\AA, $c=7.249$\,\AA. Each Y$^{3+}$ ion is surrounded by seven fluoride ions with C$_{\rm s}$ point group symmetry. The YF$_7$ polyhedra
form chains parallel to the c-axis. The distance between the intra-chain Y$^{3+}${}- Y$^{3+}$ ions is around 3.7\,\AA,
whereas the shortest inter-chain separation is \~{}5\,\AA. As with other fluoride hosts, K$_2$YF$_5$ has a {\it relatively} low Debye temperature and therefore band phonon cutoff frequency of around 480 cm$^{-1}$ \cite{raman, tuyen2020k2yf5}.

EPR has been used to investigate the ground state magnetic splittings for Ce$^{3+}$,  Gd$^{3+}$ \cite{loncke2007epr}, Er$^{3+}$, and Yb$^{3+}$ \cite{zverev2011electron}. 
Optical spectroscopy has been performed, with varying degrees of depth and rigour, on K$_2$YF$_5$ bulk single crystals doped with
Pr$^{3+}$ \cite{yin2003excitation,martin2001spectroscopic}, Nd$^{3+}$ \cite{karbowiak2012energy,yin2003spectroscopic}, Sm$^{3+}$ \cite{khaidukov2021study}, Eu$^{3+}$ \cite{jang2009luminescence}, Tb$^{3+}$ \cite{boutinaud1997effect,tuyen2020k2yf5}, Er$^{3+}$ \cite{loiko2016up, raman, peale, dean} and Tm$^{3+}$ \cite{wang2009judd,li2004spectra}.

Magnetic-splitting data has been shown to be crucial for crystal-field calculations in low-symmetry systems \cite{horvath, horvath1, test, prediction}. In those studies, directional information such as EPR-derived ground-state g-tensors and/or rotational Zeeman data were used to give a unique orientation for the axis system used in  the calculations. In high symmetries, the crystal-field parameters may be uniquely determined once an axis system is chosen. For example, in O$_\text{h}$ symmetry the four-fold axis may be chosen to be along  $x$, $y$, and $z$. In that case, the two (real) crystal-field parameters be unambiguously determined. In low symmetries it is no longer possible to unambiguously determine crystal-field parameters by fitting to electronic energy levels.  
In the case of C$_1$ (i.e.\ no) symmetry, parameters fitted to only electronic energies may be rotated arbitrarily in three dimensions, changing both the \emph{phases} and the \emph{magnitudes} of the parameters \cite{BuRe04}. 
For C$_{\rm s}$ symmetry, relevant for  K$_2$YF$_5$, the $z$ axis may be chosen perpendicular to the mirror plane. However, any rotation about $z$ leaves the electronic energies invariant and such a rotation changes the \emph{phases} of the complex parameters.

Karbowiak \emph{et al.}\  \cite{karbowiak2012energy} reported a comprehensive analysis of electronic  energy levels of Nd$^{3+}$ in  K$_2$YF$_5$. Their crystal-field fit used C$_{\rm s}$ symmetry and the starting parameters were based on the superposition model. Unfortunately, there is no magnetic-splitting data available for that system, so the orientation of calculated magnetic moments can not be checked against experiment.

EPR data is available for Er$^{3+}$-doped K$_2$YF$_5$ \cite{zverev2011electron}.
In this paper, we  present detailed spectroscopic studies of K$_2$YF$_5$ microparticles doped with 2 mol\% of Er$^{3+}$, prepared by the solvothermal method. Powder X-ray diffraction (PXRD) and scanning electron microscopy (SEM) were used to obtain information about the crystallinity and morphology of these
microcrystals.
\highlight{
The 2 mol\% doping was chosen to give a good signal in both absorption and emission. A very high concentration can quench luminescence due to cross-relaxation. 
}
A detailed energy-level scheme has been deduced from high-resolution absorption and laser excited fluorescence for samples nominally cooled to 10~K. This experimental data, supplemented by EPR data,  is used to fit a  crystal-field model using techniques that build upon our earlier work on C$_1$ sites in Y$_2$SiO$_5$ \cite{horvath1, test, prediction}. 

\section{Materials and Methods}
\subsection{Materials}

Yttrium nitrate (Y(NO$_3$)$_3$, 99.90\%) and erbium nitrate pentahydrate (Er(NO$_3$)$_3$ $\cdot$ 5H$_2$O, 99.90\%) obtained from
Sigma-Aldrich (St. Louis, USA) were used as a rare earth ion source. Anhydrous potassium fluoride (KF, BDH laboratory,
98.00\%) and potassium hydroxide (KOH, Sigma-Aldrich, 99.00\%) were used as the fluoride source. All the chemicals were
used as received. Milli-Q water, ethanol (analytical grade, 99.50\%), and Oleic acid (analytical grade, 99.50\%), were
employed to wash and prepare the samples. 

\subsection{Synthesis of K$_2$YF$_5$:Er$^{3+}$ microparticles}

K$_2$YF$_5$ microparticles doped with two mol\% Er$^{3+}$ were synthesized via the solvothermal method \cite{bian2019near,ding2015hydrothermal}. The appropriate amounts of KOH (35
mmol) were dissolved in 7.5 ml deionized water in a beaker. After that, 25 ml ethanol and 25 ml oleic acid were
added. Then a 10 mL aqueous solution containing 80 mmol of KF was added to the above solution and stirred to make a uniform
solution. Lastly, 10 ml of Y(NO$_3$)$_3$ (1.98 mmol) and Er(NO$_3$)$_3$$\cdot$5H$_2$O (0.02 mmol) aqueous solution was added to the
above-mixed solution. After stirring for 30 minutes, the as-obtained milky solution was transferred into a 100 mL of
Teflon-lined stainless-steel autoclave and then heated at 220$^\circ$C for 24 h. After that, the autoclave cooled naturally
to room temperature, and the precipitate was collected through centrifugation (7000 rpm, 10 minutes), thoroughly washed
with ethanol, and then dried at 80$^\circ$C for 12 h. 

\subsection{Characterisation and Spectroscopy}

Powder X-ray diffraction patterns were collected on a RIGAKU 3 kW SmartLab X-ray diffractometer employing a CuK$\alpha $1
radiation source, $\lambda $=1.5406 Å. Scanning electron microscope (SEM) images were obtained with a JEOL 7000F
scanning electron microscope operating at 15 kV.

The absorption measurements were carried out using an N$_{2}$ purged Bruker Vertex 80
FTIR with a resolution of 0.075 cm$^{-1}$. The powder was pressed into thin pellets using a pellet maker and then placed
into a copper sample holder, which itself was in thermal contact with the cold finger of a closed cycle cryostat, for absorption measurements. In order to perform fluorescence spectroscopy, a pulsed N$_{2}$ laser pumped PTI GL-302 dye laser was employed as the excitation source, providing wavelength tunability through the visible region. The detection system consisted of a Horiba iHR550 monochromator equipped with either a Hamatmatsu R2257P thermoelectrically cooled visible photomultiplier tube (PMT) or a thermoelectrically cooled Hamamatsu H10330C near-infrared PMT. For both absorption and fluorescence measurements a Janis CCS-150 closed-cycle cryostat cooled the sample to 10 K.

\subsection{Crystal-Field Calculations}
In our calculations, the 4f$^{11}$ configuration of Er$^{3+}$ is  modelled by a parametrized Hamiltonian \cite{carnall1989systematic,Liu2005,reid2016theory}:
\begin{equation}
    H = H_\text{FI} + H_\text{CF} + H_\text{Z}
\end{equation}
$H_\text{FI}$ is the free ion contribution,  $H_\text{CF}$ is the crystal-field contribution, and $H_\text{Z}$ is the Zeeman interaction. The crystal-field contribution is  expanded in terms of spherical tensors as:
\begin{equation}
\label{eq:cf}
    H_\text{CF} = \sum_{k,q} B_{q}^{k}C_{q}^{(k)}.
\end{equation}
The  $B_{q}^{k}$ are crystal-field parameters and the $C_{q}^{(k)}$ are  Racah spherical tensor operators acting within the 4f$^N$ configuration.
In this Hamiltonian, $k$ is restricted to 2, 4, and 6. In C$_{\rm s}$ symmetry  parameters with odd $q$ are zero, but the parameters with $q \neq 0$ are complex.
The Zeeman interaction may be written as:
\begin{equation}
 H_\text{Z} = \mathbf{B} \cdot (\mathbf{L} + g_\text{s} {S}), 
\end{equation}
where $\mathbf{B}$ is the (externally applied) magnetic field, $\mathbf{L}$ and $\mathbf{S}$ are orbital and spin angular momentum operators and $g_\text{s}$ the g-value of the electron. 

Fitting crystal-field parameters for low-symmetry systems is challenging. Our previous work on  $C_1$ sites in Y$_2$SiO$_5$ \cite{horvath1, test, prediction} has shown that the addition of magnetic-splitting data is essential to fix the orientation of the axis system. For this work, we chose the $x$, $y$ and $z$ axes to be along the $a$, $b$, and $c$ crystallographic axes. The mirror plane is, therefore, perpendicular to the $z$ axis.

\clearpage

\section{Results and Discussion}
\subsection{Phase, Morphology, and Composition}

Phase identification of the K$_2$YF$_5$: 2mol\%Er$^{3+}$ microparticles was performed using the PXRD technique. As shown in
Figures 1 (a) and (b), the PXRD patterns of the as synthesised K$_2$YF$_5$:Er microparticles are plotted against the orthorhombic K$_2$YF$_5$
crystal data (JCPDS No. 72-2387)  for comparison \cite{bian2019near,ding2015hydrothermal}. The diffraction peaks of the samples are in good agreement
with the reference data for orthorhombic K$_2$YF$_5$ (space group Pna21).
To further investigate the morphology of the K$_2$YF$_5$:Er
microparticles, scanning electron microscopy (SEM) was used. Figure 1 (c) shows that the K$_2$YF$_5$:Er sample is a mixture of
hexagonal and octagonal microcrystals with a particle size distribution ranging from 4 µm to 14 µm. 

\begin{figure}[h!]
\centering
 \includegraphics[width=\textwidth]{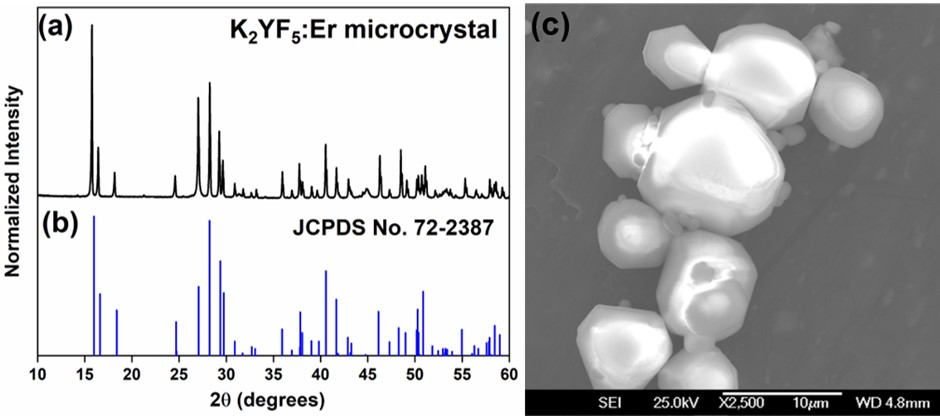} 
\caption{ \label{fig:xray}
Powder X-ray diffraction patterns of (a) the K$_2$YF$_5$: Er microcrystals and (b) the standard orthorhombic K$_2$YF$_5$ (JCPDS.
No. 72-2387) reference pattern. (c) SEM image of the K$_2$YF$_5$: Er microcrystals.}
\end{figure}

\clearpage 
\subsection{Optical Absorption and Fluorescence Spectra}

\begin{figure}
\centering
  \includegraphics[width=0.95\textwidth]{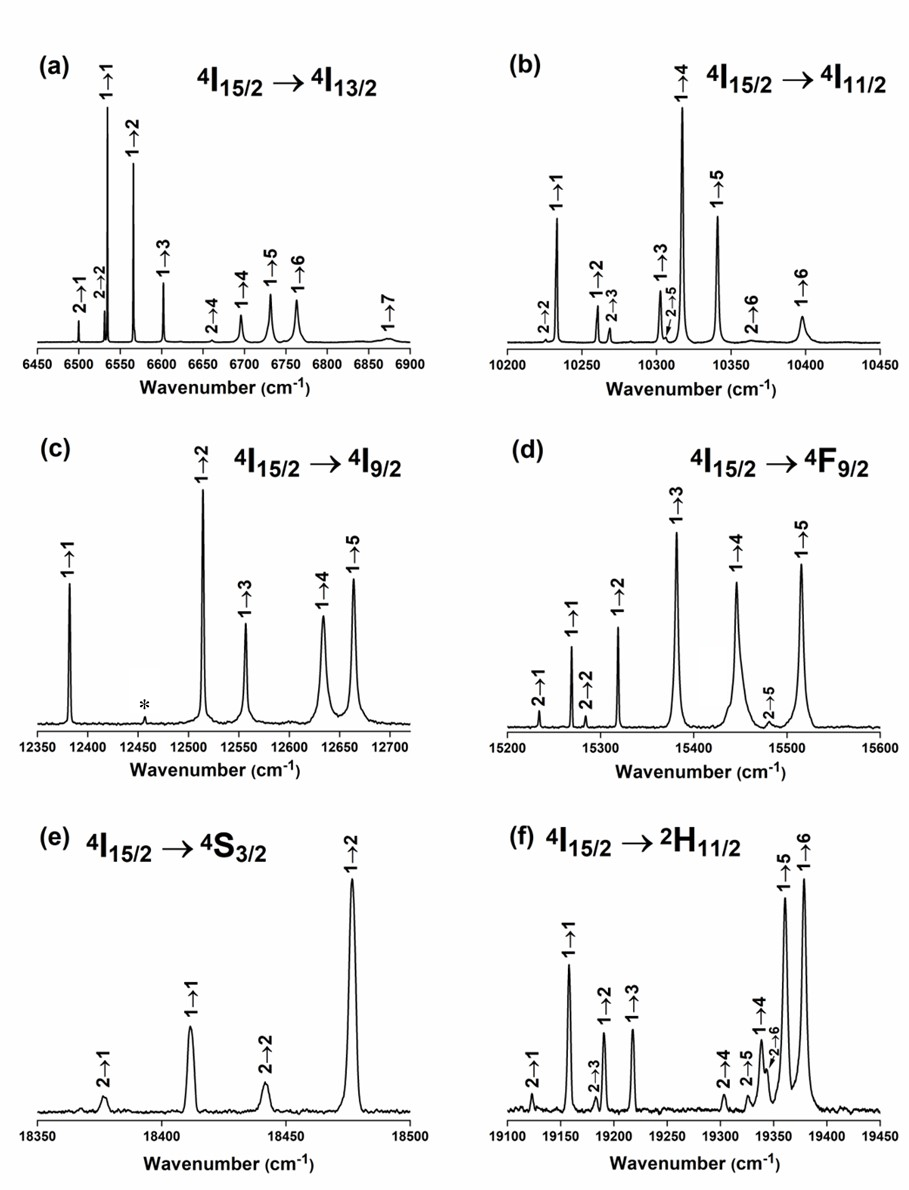} 
\caption{ \label{fig:absorption}
  10~K absorption spectra for the
(a) $^4$I$_{15/2}$ $\rightarrow$  $^4$I$_{13/2}$;
(b) $^4$I$_{15/2}$ $\rightarrow$  $^4$I$_{11/2}$;
(c) $^4$I$_{15/2}$ $\rightarrow$  $^4$I$_{9/2}$;
(d) $^4$I$_{15/2}$ $\rightarrow$  $^4$F$_{9/2}$;
(e) $^4$I$_{15/2}$ $\rightarrow$  $^4$S$_{3/2}$;
(f) $^4$I$_{15/2}$ $\rightarrow$  $^2$H$_{11/2}$
transitions of  Er$^{3+}$ doped K$_2$YF$_5$
microcrystals. The transition labelled with an asterisk is unassigned.}
\end{figure}

Figure \ref{fig:absorption} shows absorption spectra of the pelletized K$_{2}$YF$_{5}$:2\%Er$^{3+}$ microparticle ensembles for transitions between the  $^4$I$_{15/2}$ ground multiplet and the $^4$I$_{13/2}$, $^4$I$_{11/2}$, $^4$I$_{9/2}$, $^4$F$_{9/2}$, $^4$S$_{3/2}$ and $^2$H$_{11/2}$ multiplets with samples cooled to a nominal temperature of 10~K.
Numerical labels are used to indicate assignments to the intial and final states within the ground and excited multiplets. Due to the approximately 10 micron diameter of these particles, the particle size distribution does not play a substantial role in determining the measured inhomogeneous linewidths which are not significantly greater than those observed in bulk crystals \cite{dean}. One of the notable features of the spectra is the observation of transitions from thermally populated excited states within the ground multiplet. This occurs due to the difficulty in getting good thermal contact across the entire pellet with the copper sample mount. Due to the fact that even within a pelletized sample, there are limited contact points from particle to particle, the only way to fully overcome this would be to immerse the entire pellet in liquid helium. However the observation of additional transitions is, in fact, an advantage here since the purpose is to assign energy levels. All of the experimentally assigned energy levels are given in Table \ref{tab:levels}.

\begin{figure}
\centering
 \includegraphics[width=0.95\textwidth]{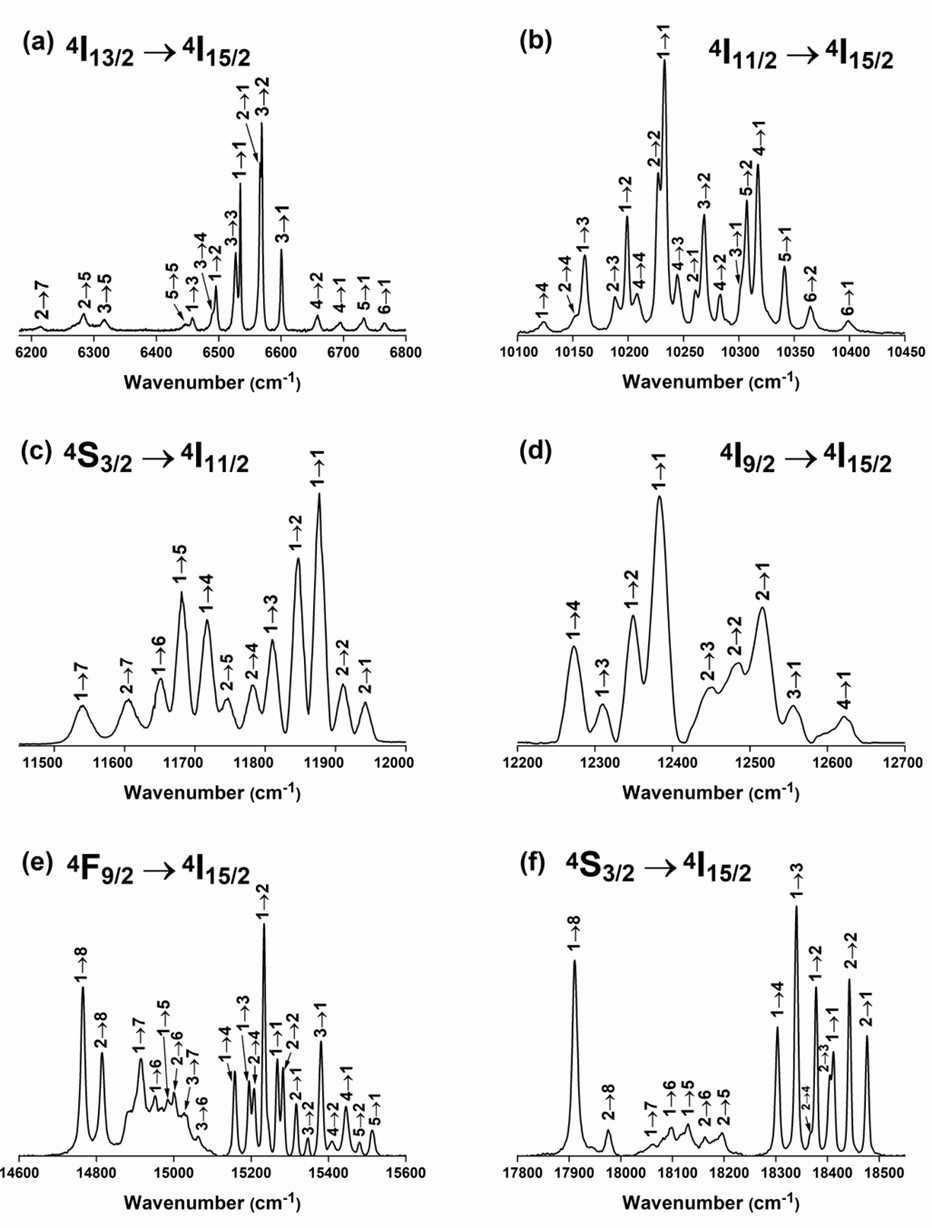} 
\caption{ \label{fig:emission}
10~K fluorescence spectra of Er$^{3+}$ doped K$_2$YF$_5$ microcrystals
\highlight{
excited at 19157 cm$^{-1}$
}
for the
transitions:
(a) $^4$I$_{13/2}$   $\rightarrow$ $^4$I$_{15/2}$;
(b) $^4$I$_{11/2}$   $\rightarrow$ $^4$I$_{15/2}$;
(c) $^4$I$_{9/2}$   $\rightarrow$ $^4$I$_{15/2}$;
(d) $^4$F$_{9/2}$    $\rightarrow$ $^4$I$_{15/2}$;
(e) $^4$S$_{3/2}$    $\rightarrow$ $^4$I$_{15/2}$;
(f) $^2$H$_{11/2}$   $\rightarrow$ $^4$I$_{15/2}$.
\highlight{
  The transitions in (c) and (d) were relatively weak, so the scans were taken with larger slit settings and wavelength steps.
}
}
\end{figure}

Fluorescence spectra were recorded throughout the visible and near infrared regions with the microparticle sample excited at 19157 cm$^{-1}$, corresponding to the $^4$I$_{15/2}$Z$_{1}$ $\rightarrow$ $^2$H$_{11/2}$F$_1$ transition. The nominal base temperature for the sample was recorded as 10~K. Figure \ref{fig:emission} shows the
\highlight{
down-shifted
}
fluorescence of the $^4$I$_{13/2}$, $^4$I$_{11/2}$, $^4$I$_{9/2}$, $^4$F$_{9/2}$, and $^4$S$_{3/2}$ multiplets to the ground   $^4$I$_{15/2}$ multiplet. As is expected for Er$^{3+}$, transitions emanating from the $^{4}$I$_{13/2}$ multiplet are amongst the strongest (and magnetic dipole allowed) whilst those from the $^{4}$I$_{9/2}$ muiltiplet are weak due to the small energy gap to the $^{4}$I$_{11/2}$ multiplet which allows for significant non-radiative depopulation. As with the absorption spectra, numerous transitions from thermally populated excited state levels can be observed for transitions from any given multiplet. The complete list of assigned crystal-field levels is presented in Table \ref{tab:levels}.

\clearpage

\clearpage

\subsection{Crystal-Field Analysis}

The parmeters in our crystal-field model are fitted to both the electronic energy levels (Table \ref{tab:levels}) and the ground-state g-tensor (Table \ref{tab:gvalues}). In our calculation, the $x$, $y$, and $z$ axes correspond to the crystallographic $a$, $b$ and $c$ axes. As discussed above, the Y$^{3+}$ sites that the Er$^{3+}$ ions substitute into have C$_{\rm s}$ symmetry, with the mirror plane being the $x$--$y$ plane \cite{loncke2007epr}. The sites have several distinct orientations, related by rotations or reflections, but there are only two sets of magnetically-inequivalent sites \cite{loncke2007epr,zverev2011electron}, with different g-tensors, as indicated in Table \ref{tab:gvalues}. The crystal-field Hamiltonian for the two sets of sites are related by a complex conjugtion of the crystal-field parameters.

The fitting approach followed Ref.\ \cite{horvath1}.
Calculations of zero-field electronic energy levels were carried out in parallel to calculations for  12 orientations of the magnetic field. The weighted $\chi^2$ used in the minimization process consisted of squares of differences between the calculated and experimental electronic energies energies and squares of differences between the ground-state magnetic splittings and splittings calculated from the g-tensor. The magnetic splittings were weighted more heavily than the electronic energies. 
A basin-hopping algorithm was used to search the parameter space. We allowed the Coulomb, $F^k$, spin-orbit, $\zeta$, and crystal-field, $B_q^k$, parameters to vary freely. Other parameters  were fixed at the fitted values reported by Carnall \emph{et al.}\  for Er$^{3+}$ in LaF$_3$ \cite{carnall1989systematic}.
An intemediate-coupled basis set was used, trucated at 30,000\,cm$^{-1}$ (15 $J$ multiplets), to speed up the calculation.

Table \ref{tab:crystalfield} gives fitted parameters for Nd$^{3+}$ in K$_2$YF$_5$ \cite{karbowiak2012energy} and for two different fits for  Er$^{3+}$ in K$_2$YF$_5$. Fit A was obtained by using the  Nd$^{3+}$ parameters as a starting point. This gave a good fit but we also searched for a global minimum by trying other starting points. The best fit we obtained is presented as Fit B. Uncertainties for this fit were estimated using the Monte-Carlo approach described in Ref.\  \cite{horvath1} and the predicted energy levels for these parameters are given in Table  \ref{tab:levels}.
For both fits the calculated g-tensor agrees with the experimental data (Table \ref{tab:gvalues}) to the accuracy of the measurements.
For Fit A, the signs of  $B^2_0$, $B^4_0$,  and  $B^6_0$ are the same as for Nd$^{3+}$. Note that since the  Nd$^{3+}$ fit did not include any magnetic splitting data, the overall phase of the complex parameters is undetermined.  Fit B  is quite different, with opposite signs from the  Nd$^{3+}$ fit for  $B^4_0$   and  $B^6_0$.

Karbowiak \emph{et al.}\ \cite{karbowiak2012energy} used the superposition model to investigate the relationship between the Nd$^{3+}$ crystal-field parameters and the geometry of the site. In the superposition-model calculations, they assumed 
the  $Pna2_1$ space group and therefore C$_1$ site symmetry, whereas the EPR results of Refs.\ \cite{loncke2007epr,zverev2011electron} are consistent with the $Pnam$ space group and C$_{\rm s}$ site symmetry. However, the geometry they determined had approximate C$_{\rm s}$ symmetry and that symmetry was used in their crystal-field calculation. 
Their calculation (Table 5 of \cite{karbowiak2012energy}) predicts a positive $B^2_0$, a positive  $B^4_0$, and a negative  $B^6_0$. Their Nd$^{3+}$ fit gives a negative $B^2_0$, a positive  $B^4_0$, and a negative  $B^6_0$. However, as they point out, the $k=2$ parameters are expected to be more sensitive to long-range electrostatic contributions than the $k=4$ and $k=6$ parameters. 
Our best fit (Fit B) has opposite signs to the superposition-model prediction for all three of  $B^2_0$, $B^4_0$,  and  $B^6_0$, whereas our Fit A, like the  Nd$^{3+}$ fit, is in agreement for $B^4_0$,  and  $B^6_0$. Our energy-level data set is relatively small, so it is possible that the inclusion of more energy-level or magnetic-splitting data would result in a global minimum in better agreement with the superposition model.

\clearpage

\begin{table}[tb!]
\caption{ \label{tab:levels}
Experimental and fitted electronic energy levels for 
 Er$^{3+}$ doped K$_2$YF$_5$ microcrystals. All energies are in cm$^{-1}$. The experimentally determined energy levels have an
 associated uncertainty of 1\,cm$^{-1}$.
The calculated levels are from the Fit B parameter set of Table \ref{tab:crystalfield}. 
 The alphanumeric labels follow the convention of Dieke \cite{dieke}.}

\footnotesize

\begin{tabular}{ccr@{\hskip 10mm}r}
\hline
Multiplet         &State             & Measured            &Fit \\
\hline

$^4$I$_{15/2}$     &Z$_1$             &            0                   &          $-2 $          \\
                  &Z$_2$             &            35                  &           41         \\
                  &Z$_3$             &            74                  &           77 
  \\
                  &Z$_4$             &           111                  &          111         \\
                  &Z$_5$             &           283                  &          279         \\
                  &Z$_6$             &           314                  &          314         \\
                  &Z$_7$             &           349                  &          357         \\
                  &Z$_8$             &           502                  &          489         \\ \\
$^4$I$_{13/2}$     &Y$_1$             &         6534                   &        6539          \\
                  &Y$_2$             &          6566                  &         6568         \\
                  &Y$_3$             &          6601                  &         6602         \\
                  &Y$_4$             &          6695                  &         6697         \\
                  &Y$_5$             &          6731                  &         6739         \\
                  &Y$_6$             &          6763                  &         6761         \\
                  &Y$_7$             &          6873                  &         6873         \\ \\
$^4$I$_{11/2}$     &A$_1$             &        10233                   &       10222          \\
                  &A$_2$             &         10260                  &        10250         \\
                  &A$_3$             &         10302                  &        10297         \\
                  &A$_4$             &         10317                  &        10316         \\
                  &A$_5$             &         10340                  &        10341         \\
                  &A$_6$             &         10397                  &        10407         \\ \\
$^4$I$_{9/2}$     &B$_1$             &         12382                  &        12380         \\
                  &B$_2$             &         12514                  &        12506         \\
                  &B$_3$             &         12556                  &        12559         \\
                  &B$_4$             &         12633                  &        12626         \\
                  &B$_5$             &         12664                  &        12677         \\ \\
$^4$I$_{9/2}$     &D$_1$             &         15268                  &        15272         \\
                  &D$_2$             &         15318                  &        15313         \\
                  &D$_3$             &         15381                  &        15377         \\
                  &D$_4$             &         15445                  &        15446         \\
                  &D$_5$             &         15515                  &        15522         \\ \\
$^4$S$_{3/2}$      &E$_1$             &        18411                   &       18403          \\
                  &E$_2$             &         18476                  &        18476         \\ \\
$^2$H$_{11/2}$     &F$_1$             &        19157                   &       19166          \\
                  &F$_2$             &         19190                  &        19199         \\
                  &F$_3$             &         19217                  &        19213         \\
                  &F$_4$             &         19338                  &        19333         \\
                  &F$_5$             &         19360                  &        19363         \\
                  &F$_6$             &         19378                  &        19375         \\ \\
  
\hline
\end{tabular}

\end{table}

\clearpage 

\begin{table}[tb!]
\caption{ \label{tab:gvalues}
  Experimental ground-state g-tensors for the two magnetically-inequivalent sites in Er$^{3+}$ doped K$_2$YF$_5$ in the crystal axis system ($x=a$, $y=b$, $z=c$).
  Principal values from Ref.\ \cite{zverev2011electron} are
  $g_x = 0.81$, $g_y=13.6$, $g_z = 3.19$, with $z$ along the crystal $c$ axis and $x$ and $y$ in the $a$-$b$ plane with $x$ rotated from $a$ by an angle of $\pm 87.6^\circ$. 
  }
\footnotesize
\renewcommand{\arraystretch}{1.3}
\begin{tabular}{cccccc}
\hline
                      $g_{xx}$     &   $g_{xy}$     &   $g_{xz}$   & $g_{yy}$   & $g_{yz}$   & $g_{zz}$ \\
\hline
                     $13.59$      &   $\mp 0.54$      &   $0$        & $0.83$    & $0$        & $3.19$ \\

  \hline
\end{tabular}
\end{table}

\begin{table}[tb!]
\caption{ \label{tab:crystalfield}
  Crystal-field parameters for Nd$^{3+}$ and Er$^{3+}$ doped K$_2$YF$_5$ in units of cm$^{-1}$. 
  In Fit A and Fit B the free-ion  parameters not given in the Table were fixed to the values obtained for  Er$^{3+}$ doped LaF$_3$ \cite{carnall1989systematic}.
  The Nd$^{3+}$ parameters are from Ref.\ \cite{karbowiak2012energy}. Fit A was obtained by using the  Nd$^{3+}$  parameters as a starting point. Fit B was the best fit obtained from a global search. 
  Weighted $\chi^2$ for the Fit A and Fit B  were 5.39 and  4.37 resectively. 
}
\footnotesize
\renewcommand{\arraystretch}{1.3}
\begin{tabular}{cc@{\hskip 10mm}c@{\hskip 10mm}c@{\hskip 10mm}c}
\hline
Parameter       & Nd$^{3+}$ \cite{karbowiak2012energy}& Er$^{3+}$ Fit A     &          Er$^{3+}$  Fit B            &        Uncertainty   \\
\hline                                                                
                                                                        
$E_\text{avg}$ &                      &$          35631        $&$             35630          $&$      27 	                  $    \\
$F^2$         &                      &$           96310        $&$            96290          $&$      35  	                  $     \\
$F^4$         &                      &$           68451        $&$            68479          $&$      45	                  $   \\         
$F^6$         &                      &$           53383        $&$            53371          $&$      19	                  $     \\
$\zeta$       &                      &$            2361        $&$             2361          $&$      16 	                  $     \\ \\ 
$B^2_0$       &   $-585$             &$            -256        $&$             -379          $&$      43	                  $     \\ 
$B^2_2$       &    $322 - 63i$       &$    -348   -195   i     $&$      -89   -408   i       $&$      47 +  62 i                  $   \\
$B^4_0$       &   $77$               &$             792        $&$            -1267          $&$      14	                  $     \\
$B^4_2$       &    $-1069+1167i$     &$    1220   +580   i     $&$       -72   -65   i       $&$      34 +  40 i                  $   \\
$B^4_4$       &    $353-108i$        &$     143   -666   i     $&$    -1294   -142   i       $&$      31 +  21 i                  $   \\
$B^6_0$       &   $-521$             &$            -544        $&$              270          $&$      15	                  $     \\
$B^6_2$       &    $197 + 19i$       &$       41   -75   i     $&$       134   -85   i       $&$      42 +  18 i                  $    \\
$B^6_4$       &    $-428-492$        &$      -5   +162   i     $&$     -340   -102   i       $&$      24 +  31 i                  $    \\
$B^6_6$       &    $385-470i$        &$     -284   +18   i     $&$     -192   +186   i       $&$      27 +  40 i                  $    \\

\hline
\end{tabular}
\end{table}
  
\clearpage

\section{Conclusions}

We have presented detailed optical spectroscopy of Er$^{3+}$ in K$_2$YF$_5$ microcrystals and deduced energy levels up to 20,000\, cm$^{-1}$. The optical data has been combined with EPR data for the magnetic splitting of the ground state to obtain a crystal-field fit with a known axis orientation.
Crystal-field modelling provides a powerful tool to understand the electronic structure of lanthanide ions in crystal hosts and hence to optimise applications.  We have shown that in low-symmetry hosts the addition of magnetic measurements provides crucial information for such modelling.

\printcredits



\end{document}